\documentstyle[11pt,newpasp,twoside,astrobib,psfig]{article}
\def\edcomment#1{\iffalse\marginpar{\raggedright\sl#1\/}\else\relax\fi}
\reversemarginpar

\begin{document}
\title{Radio-Variability in Radio-Quiet Quasars and Low-Luminosity AGN} 

\author{Heino Falcke}
\affil{Max-Planck-Institut f\"ur Radioastronomie, Auf dem H\"ugel 69,
53121~Bonn, Germany \\(hfalcke@mpifr-bonn.mpg.de)}
\author{Joseph Leh\'ar}
\affil{Harvard-Smithsonian Center for Astrophysics, 60 Garden Street, Cambridge, MA 02138}
\author{Richard Barvainis}
\affil{National Science Foundation, 4201 Wilson Boulevard, Arlington, VA~22230 \\(rbarvai@nsf.gov)}
\author{Neil M. Nagar\altaffilmark{1},  Andrew S. Wilson\altaffilmark{2}}
\affil{Astronomy Department, University of Maryland, College Park,
MD~20742-2421 \\(neil,wilson@astro.umd.edu)}
\altaffiltext{1}{Present Address: Osservatorio Astrofisico di Arcetri, 
Largo E. Fermi 5, I-50125 Florence, Italy}
\altaffiltext{2}{Adjunct Astronomer, Space Telescope Science Institute}

\begin{abstract}
We report on two surveys of radio-weak AGN to look for radio
variability. We find significant variability with an RMS of 10-20\% on
a timescale of months in radio-quiet and radio-intermediate
quasars. This exceeds the variability of radio cores in radio-loud
quasars (excluding blazars), which vary only on a few percent
level. The variability in radio-quiet quasars confirms that the radio
emission in these sources is indeed related to the AGN.  The most
extremely variable source is the radio-intermediate quasar III~Zw~2
which was recently found to contain a relativistic jet.

In addition we find large amplitude variabilities (up to 300\%
peak-to-peak) in a sample of nearby low-luminosity AGN, Liners and
dwarf-Seyferts, on a timescale of 1.5 years. The variability could be
related to the activity of nuclear jets responding to changing
accretion rates. Simultaneous radio/optical/X-ray monitoring also for
radio-weak AGN, and not just for blazars, is therefore a potentially
powerful tool to study the link between jets and accretion flows.
\end{abstract}

\section{Introduction}
In the past a lot of emphasis has been put on studying the radio
variability of radio-loud AGN and specifically those of blazars
\cite{WagnerWitzel1995}. There the radio emission is most certainly 
due to a relativistically beamed jet and one goal of multi-wavelength
monitoring, including radio, is to understand particle acceleration
processes in the jet plasma as well as the relativistic effects
associated with the changing geometry and structure of jets.

On the other hand, for radio-weak AGN -- here meant to include
everything but radio-loud quasars -- the situation is somewhat
different and the database is much sparser. In fact, very few surveys
exist that address the issue of radio variability in either
radio-quiet quasars or low-luminosity AGN such as Liners and
dwarf-Seyferts (e.g.,
\citeNP{BarvainisLonsdaleAntonucci1996,HovanDykPooley1999}). In many
of these cases we are not even entirely sure that the radio emission
is indeed related to the AGN itself.

It has been proposed that radio jets are a natural product of AGN,
even that accretion flow and jet form a symbiotic system
\cite{FalckeBiermann1995}, and this view seems to catch on (e.g.,
\citeNP{Livio1997}). This  also implies a prediction for radio emission
from nuclear jets across the astrophysical spectrum of AGN, including
those being of low power or being radio-weak
\cite{FalckeBiermann1999}. For radio-quiet quasars some evidence
exists that this is indeed the case, like the finding of optical/radio
correlations (e.g.,
\citeNP{BaumHeckman1989,MillerRawlingsSaunders1993,FalckeMalkanBiermann1995}), 
or the detection of high-brightness temperature radio cores in a few
radio-quiet quasars \cite{BlundellBeasley1998}.

Clearly, if (some of) the radio-emission in radio-weak AGN is coming
from the central engine, we would expect to see a certain degree of
radio variability as seen in other wavebands. Finding this would,
firstly, confirm the AGN nature of the radio emission and, secondly,
allow us to study the link between accretion flows and radio jet in
more detail. In a symbiotic picture of accretion disk and radio jet
one would expect to see a change in the accretion rate first reflected
in a change in optical emission and then later in a change in the
radio emission. The type of radio variability found in radio-weak AGN
should also depend on whether or not the jets are relativistic and
whether or not they are pointing towards the observer.

To start addressing some of these questions we have started a number
of projects targeted at different classes of AGN -- mainly radio-quiet
quasars and LINERs. In the following we will present a report of first
and preliminary results of these projects.

\section{Radio-Quiet and Radio-Intermediate Quasars}
To study the radio-variability of quasars we selected a sample of
thirty sources from the PG quasar sample
\cite{SchmidtGreen1983,KellermannSramekSchmidt1994}, the LBQS sample
\cite{VisnovskyImpeyFoltz1992,HooperImpeyFoltz1995,HooperImpeyFoltz1996}, 
and the NVSS \cite{BischofBecker1997}. The sources were selected to
give a detectable flux density at 3.6 cm (8.5 GHz) above 0.3 mJy and
to roughly equally fill the parameter space of the radio-to-optical
flux ratio ($R$), including radio-quiet (RQQ, $R<3$),
radio-intermediate (RIQ, $3<R<100$), and radio-loud quasars (RLQ,
$R>100$, see
\citeNP{FalckeSherwoodPatnaik1996}). In the end we had 10, 13, and 7
objects respectively in each category.

The quasars were observed with the VLA roughly every month for eight
epochs and then at one more epoch a year later. Integration times
varied between 2 and 12 minutes. Where applicable, i.e. for some
radio-loud quasars, we picked out the compact core and ignored
emission from the extended lobes. The other sources appeared
point-like on the maps.

\begin{figure}[h]\label{lcurve}
\centerline{\psfig{figure=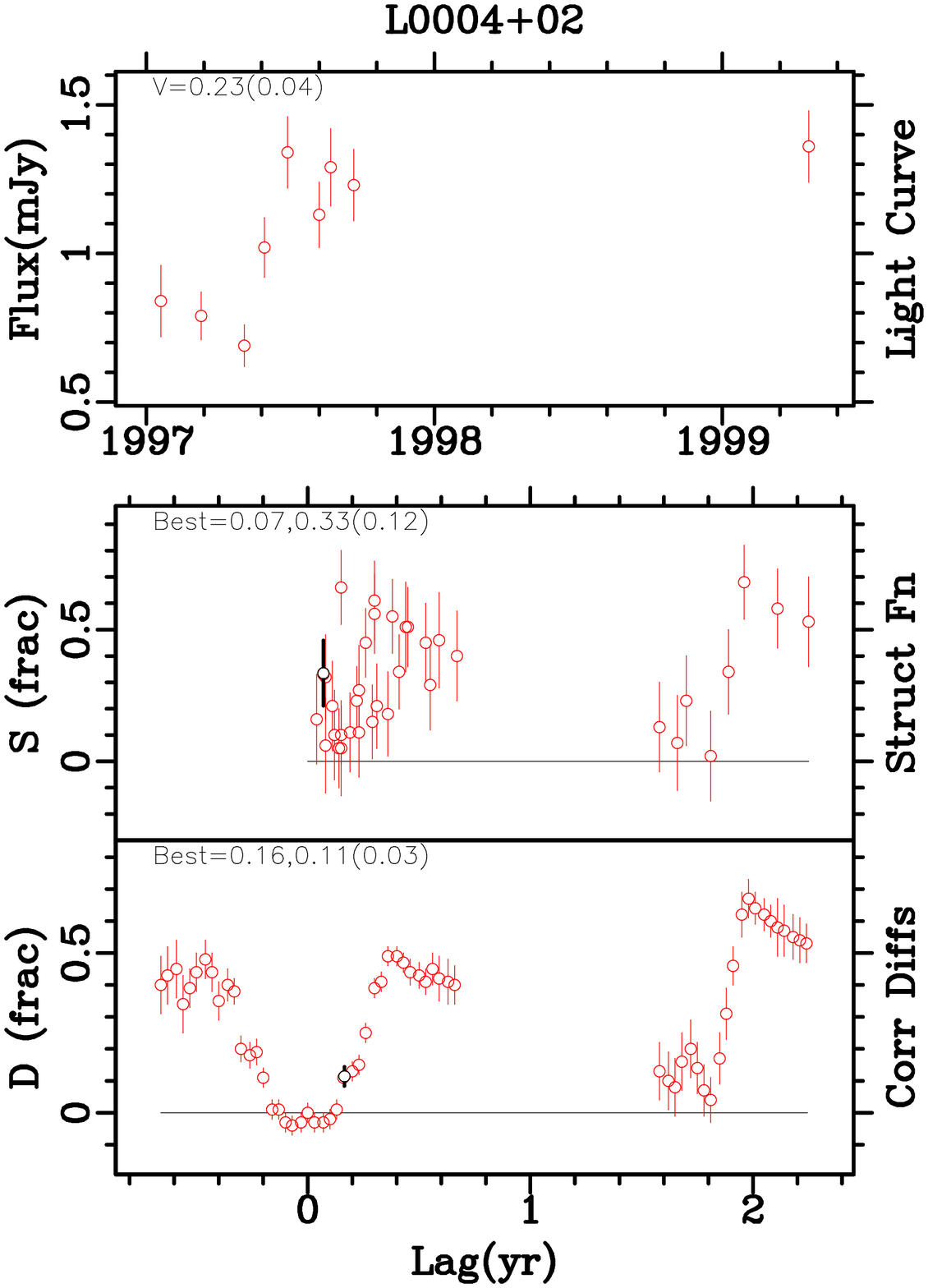,height=0.17\textheight,bblly=17.1cm,bbury=25.7cm,bbllx=1.5cm,bburx=18.2cm,clip=}~\psfig{figure=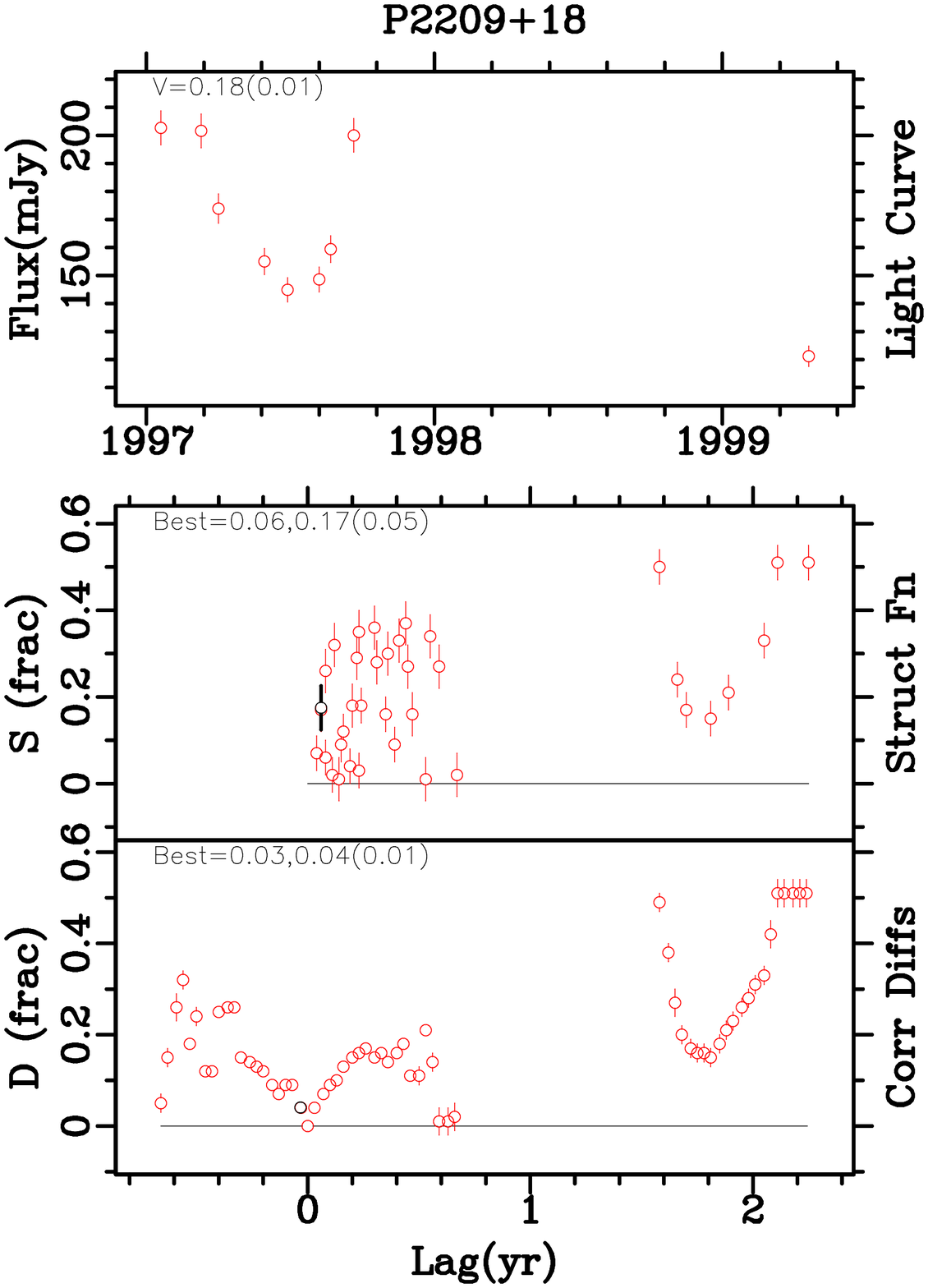,height=0.17\textheight,bblly=17.1cm,bbury=25.7cm,bbllx=2.7cm,bburx=18.2cm,clip=}}\vspace{3mm}
\centerline{\psfig{figure=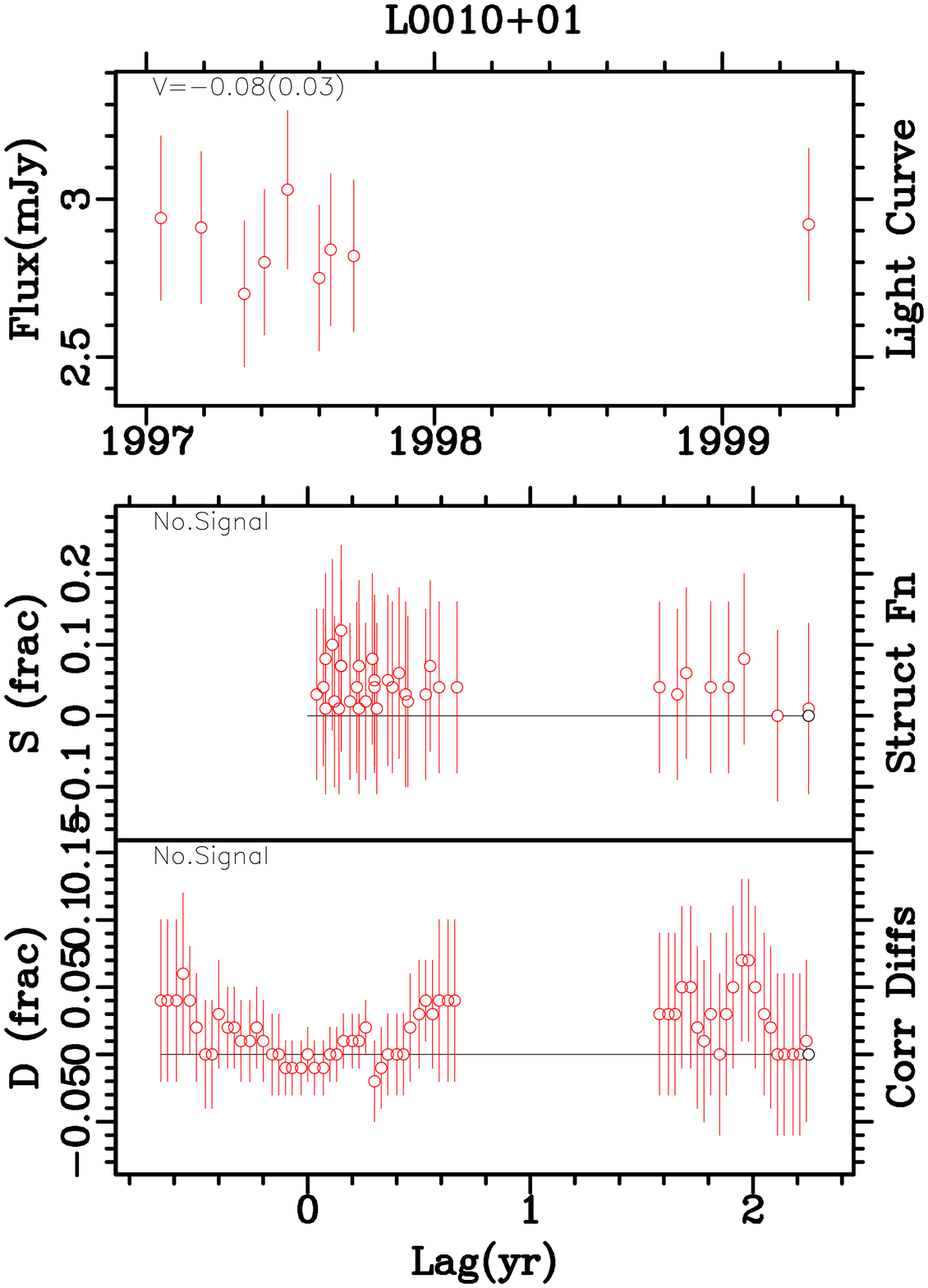,height=0.17\textheight,bblly=17.1cm,bbury=25.7cm,bbllx=1.5cm,bburx=18.2cm,clip=}~\psfig{figure=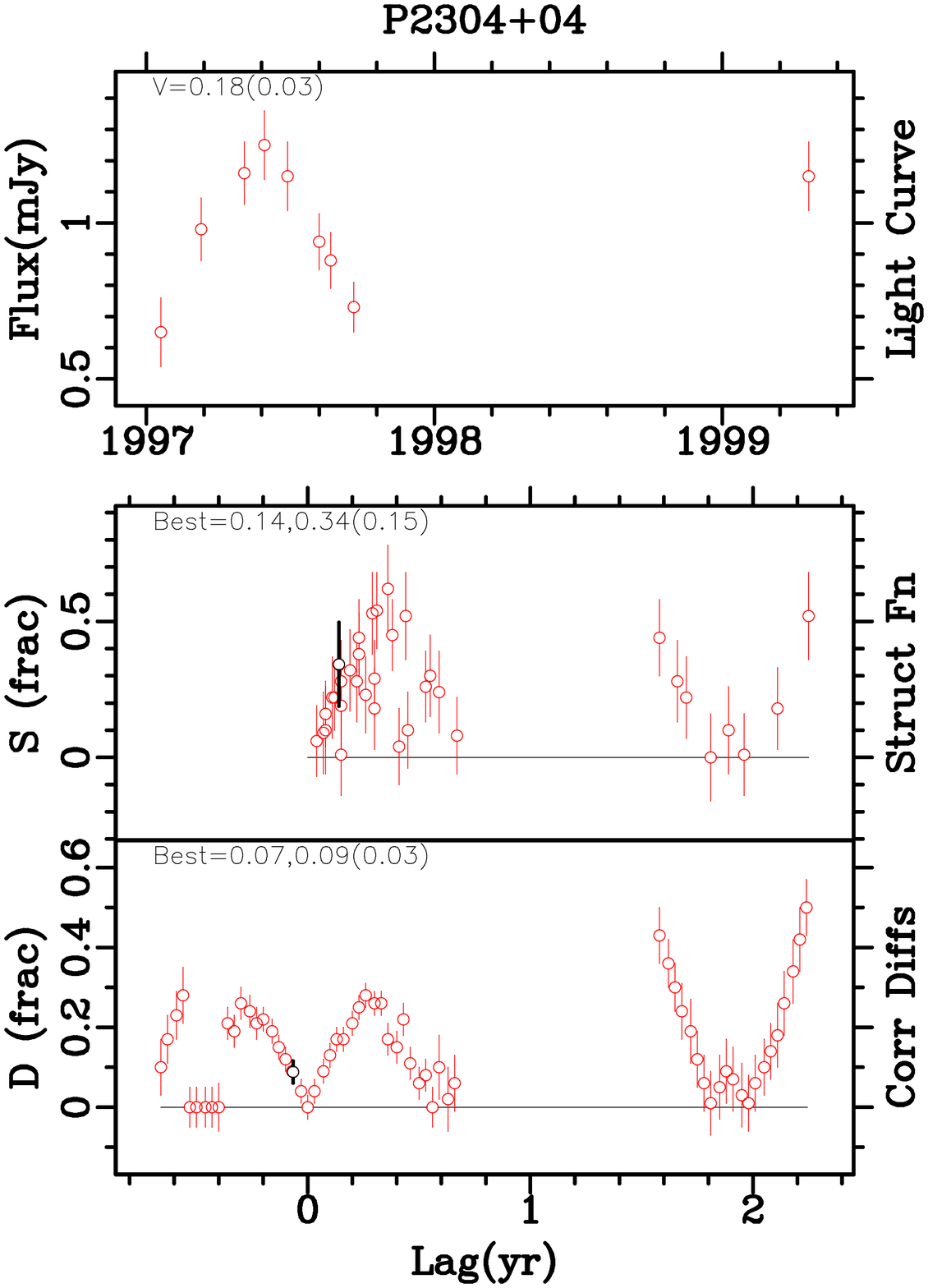,height=0.17\textheight,bblly=17.1cm,bbury=25.7cm,bbllx=2.7cm,bburx=18.2cm,clip=}}
\caption[]{Light Curves of four selected quasars -- three radio-quiet and
one radio-intermediate -- from our survey as observed with the VLA at
8.5 GHz over two years. Three of the sources show significant
variation within a year, while L0010+01 does not.}
\end{figure}

A few sample light curves are shown in Figure~1. Error bars
include statistical and systematic (calibration uncertainties)
errors. The figure shows that in some cases we have distinct flux
density variations within one year. Despite the rather low,
i.e. milli-Jansky, flux density level we are able to clearly trace the
variations from month to month in some of these galaxies. For
comparison we also show one rather faint quasar where we consistently
measure a constant flux density from epoch to epoch. This demonstrates
that measuring radio-variability even in radio quiet quasars is not
too much of a daunting task anymore.

\begin{figure}[h]\label{rqqvar}
\centerline{\psfig{figure=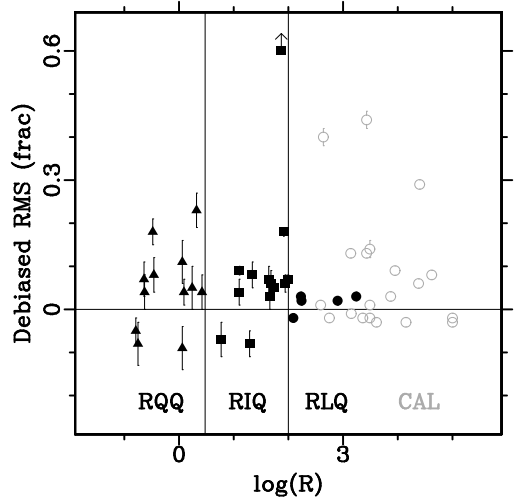,width=0.5\textwidth,bblly=5.1cm,bbllx=1.7cm,bbury=21.4cm,bburx=18cm}}
\caption[]{Debiased variability index for all sources in our
survey versus the radio-to-optical flux ratio ($R$-parameter). A
negative index indicates a non-significant variability.  The sources
are categorized simply as radio-quiet (triangles), radio-intermediate
(boxes), and radio-loud (squares) according to their $R$-parameter
alone.  Open circles represent calibrator sources many of which are
typically blazars. The variability index of III~Zw~2 -- the highest
point in the diagram -- is shown here only as a lower limit to keep
the scale of the plot in reasonable bounds.}
\end{figure}

The overall result of our survey is shown in Fig.~2, where we plot a
debiased variability index $V$ against the $R$-parameter. The index is
defined here as
\begin{equation}
V={\sqrt{\Sigma\left(S_\nu(t)-\left<S_\nu\right>\right)^2 -
\Sigma\sigma^2}\over N\cdot\left<S_\nu\right>},\end{equation}
where $N$ is the number of data points, $\sigma$ is the measurement error,
and $\left<S_\nu\right>$ is the mean flux density (see
\citeNP{AkritasBershady1996}). We set the index to be negative when
the value inside the square root becomes negative (i.e., for
non-variable sources where the error bars are too conservative).

In about 80\% of the sources we find at least some marginal evidence
for variability. The variability index is about 10-20\% in the RQQs
and RIQs and only a few percent for RLQs. Most of the radio cores in
the RIQs and RLQs have flat to inverted radio spectra and there may be
a trend for higher variability with more inverted spectra.

We point out, that our sample does not include blazars. However, many
of our phase calibrators naturally are. Surprisingly, these
heterogeneously selected calibrators do show a variability that is not
too distinct from the RLQs \& RIQs in our sample.

The nature of RIQs had been discussed in the literature
before. \citeN{MillerRawlingsSaunders1993} and
\citeN{FalckeSherwoodPatnaik1996} had suggested that they could be the
relativistically boosted counter-parts to radio-quiet quasars. And
indeed three out of the two RIQs discussed by
\citeN{FalckeSherwoodPatnaik1996}, III Zw 2 and PG2209+18, are
included here and show some of the highest variability amplitudes
observed in our survey. Recently \citeN{BrunthalerFalckeBower2000}
detected superluminal expansion -- a clear indication of relativistic
motion -- in the former\footnote{However, boosting seems not to be the
sole explanation for the enhanced radio flux in III~Zw~2, strong
interaction of the relativistic jet with the ISM or torus on the
sub-parsec scale also seems to be important.}. The fact that we find
similarly strong variability in some RQQs could also point to the
activity of relativistic jets.  Clearly, since the radio emission at
centimeter wavelengths should come from the parsec scale (because of
self-absorption arguments, e.g.,
\citeNP{FalckeBiermann1995}) -- a variability timescale of months
could not be achieved by jets with highly sub-luminal speeds.

Overall, the finding of variability in many RQQs and RIQs strengthens
the conclusion that the radio emission detected in these quasars is
indeed produced by the AGN. The rather low level of variability in the
cores of radio-loud quasars is rather puzzling and might be related to
larger black hole masses and thus longer timescales. The absence of
relativistic beaming due to larger inclination angles (in contrast to
blazars) and perhaps the presence of slow-moving cocoons surrounding
the inner fast jets (e.g., Cygnus A, \citeNP{KrichbaumAlefWitzel1998})
could also play a role.

\section{LLAGN: LINERs and Dwarf-Seyferts}
Another group of AGN for which radio variability has not been studied
in a coherent fashion are low-luminosity AGN (LLAGN). Almost a third of
all galaxies in our cosmic neighborhood show evidence for low-level
nuclear activity in emission lines, i.e. show Liner or Seyfert spectra
\cite{HoFilippenkoSargent1997b}. In many of these cases it is not
entirely clear whether the activity is due to stars or a central black
hole.

We have used the VLA and VLBA to observe two samples of nearby
LLAGN. One of them was a distance-limited sample of LLAGN within 19
Mpc, the other consisted of a collection of 48 well-studied Liners and
a few dwarf-Seyferts. The VLA survey revealed a remarkable high
detection rate of compact radio cores at 15 GHz
\cite{NagarFalckeWilson2000}. Initial VLBA observations of the smaller
sample confirm that these sources have high brightness temperature
radio cores indicative of AGN \cite{FalckeNagarWilson2000}. For the
sources in our combined samples with flux densities above 3 mJy at 15
GHz and a flat radio spectrum we have now a 100\% detection rate with
VLBI \cite{FalckeNagarWilson2000b}.  This shows that a large fraction
($\sim$50\%) of LINERS and dwarf-Seyferts are indeed genuine AGN. In
addition, having two frequencies and in some cases more, we find no
evidence for highly inverted radio cores as predicted in the ADAF
model: the (non-simultaneous) spectral indices are on average around
$\alpha=0.0$.  In the six brightest sources we detect extended
emission which appears to originate in jets. Together with the
spectral indices this suggests that the nuclear emission at centimeter
radio waves is largely dominated by emission from radio jets rather
than an ADAF \cite{FalckeBiermann1999}, very similar to the situation
in more luminous AGN. The energy released in these jets could be a
significant fraction of the energy budget in the accretion flow.
Hence, there is ample  reason to also consider the radio variability
of LLAGN and perhaps learn more about the underlying black
hole/accretion flow system powering them.

As a by-product of our observing program, we have a number of sources
that were observed several times and hence can be used to obtain some
initial and basic information on LLAGN radio variability. In fact, all
18 sources in our combined samples with flux densities above 3 mJy
(the sample studied also by the VLBA) were observed at least two
times; those who were part of our first sample (48 LLAGN) were
observed three times, all epochs separated by roughly 1.5 years. All
observations were made in a similar manner at 15 GHz with the VLA in
A-configuration. This way we are not affected by resolution effects --
from the VLBI observations we know that basically all the flux on this
scale and at this frequency comes from a compact mas-component,
i.e. the core.

\begin{figure}[h]\label{llagnvar1}
\centerline{\psfig{figure=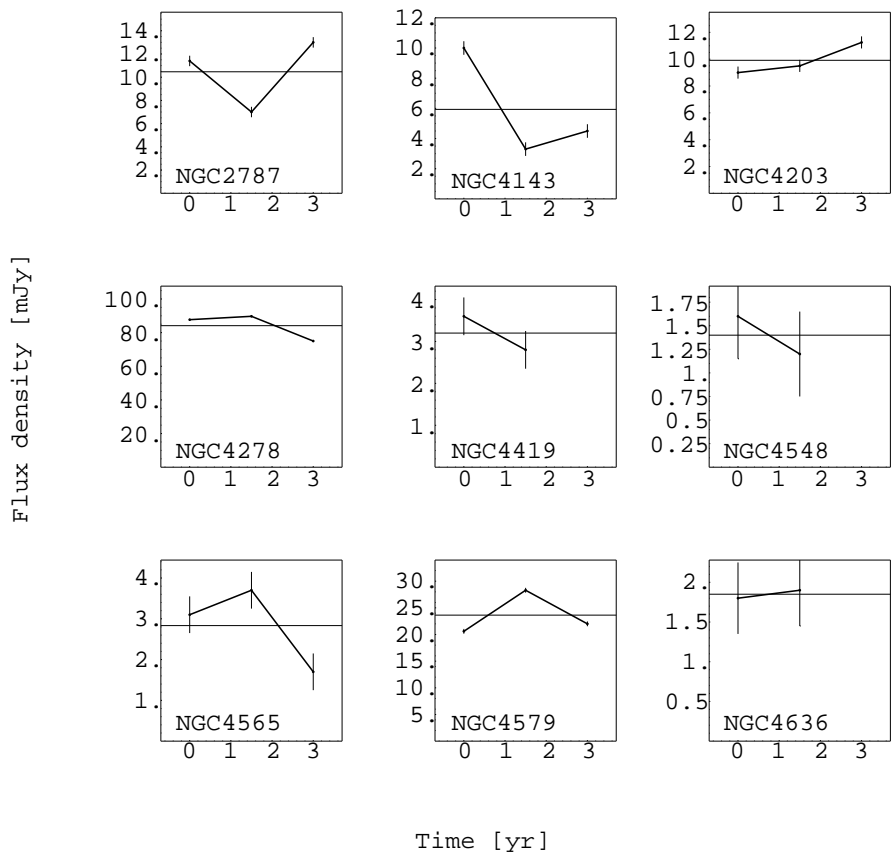,width=0.9\textwidth}}
\caption[]{Radio light curves at 15 GHz taken with the VLA in
A-configuration for radio cores in low-luminosity AGN from our
sample. The vertical line gives the average flux density level for all
epochs. The error bars only reflect the r.m.s. error and not
calibration uncertainties.}
\end{figure}

As an example, simple light curves are shown in
Figure~3. Again, the milli-Jansky level radio flux density
is not a major problem in detecting variability. Surprisingly, we find
a number of sources with rather large variations. Highly significant
peak-to-peak variability of 200-300\% is seen for example in the radio
cores of NGC2787, NGC4143, and NGC4565.

\begin{figure}[h]\label{histovar}
\centerline{\psfig{figure=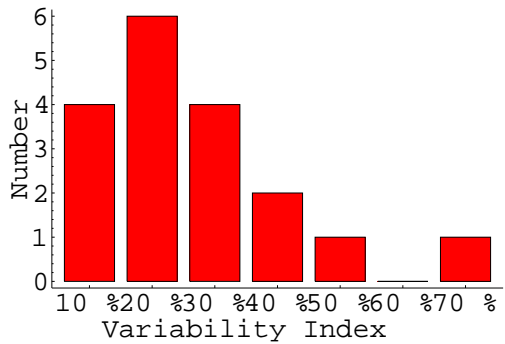,width=0.5\textwidth}}
\caption[]{Distribution of variability index for all LLAGN in our
samples with flux density larger than 3 mJy at 15 GHz. This is very
preliminary, since it includes only 2-3 epochs per source.}
\end{figure}

For such sparsely sampled light curves a variability index is rather
ill-defined for individual sources. Nevertheless, we can assume that
in a statistically useful sample as ours the distribution of the
variability index (i.e. the r.m.s. divided by the mean), as shown in
Fig.~4, will have some meaning. The general trend of this distribution
confirms the first impression from looking at the light curves:
variability on a timescale of years is common place among LLAGN and
amplitudes can reach rather large values -- from 20-70\%.

This variability is even larger than the one seen in quasars. The
rather large fraction of LLAGN with radio cores and the fact that our
sample was initially optically selected, speaks for a rather broad
range of inclination angles. Strong variations in the accretion rate
rather than effects of relativistic boosting therefore seem to be a
more likely explanation for the variability. This would be in line
with some of the X-ray variability seen in LLAGN where on scales of
years the flux has changed by factors of a few (e.g.,
\citeNP{UttleyMcHardyPapadakis1999}).

The apparent difference in variability index between LLAGN and RQQs
seen here could be related to possibly smaller black hole masses in
the former. Since we are probing only a narrow range of time scales in
our programs it could well be that for larger black hole masses the
time scale of strong variability is significant larger than a year and
hence remains undetected. Alternatively, one could postulate a
different type of accretion which is more volatile in LLAGN than in
quasars. For example, if the accretion onto the central black hole is
fed by stellar winds from a few sources only, as speculated for
example for the Galactic Center \cite{CokerMelia1997}, then evolution
and change in orbits of individual stars can have a much more
pronounced effect on the overall accretion rate than in a situation
where the accretion proceeds through a large scale and massive
accretion disk.

\section{Summary}
We have established significant intra-year variability in a sample of
radio-quiet and radio-intermediate quasars as well as in a sample of
low-luminosity AGN. The variability in quasars strengthens the notion
that also in supposedly radio-quiet quasars the radio emission is
produced by the AGN -- a large fraction of that very close to the
central engine. The strong variability could be related to the
presence of relativistic jets in at least some RQQs and RIQs.

In the radio cores of low-luminosity AGN the radio variability on the
time\-scales probed here -- roughly one year -- seems to be even
higher. Lower black hole masses could be one possible explanation.

The detection of radio cores and radio-variability in these sources
opens up the possibility to obtain a closer look on the connection
between jet formation and accretion flows through coordinated
optical/X-ray/radio monitoring also for radio-weak AGN. Changes in the
accretion rate should be reflected also in the radio emission. Already
now one can speculate that from the large radio-variability of some
Liners and dwarf-Seyfert rather large fluctuations in the accretion
rate are expected. Future long-term monitoring campaigns should
therefore seriously consider including radio monitoring as well, even
if the flux densities are only a few milli-Jansky in a compact core.

\acknowledgments this works summarizes two partially unpublished
results from various projects. The work was split in the following
manner: RB \& JL were involved in the quasar monitoring while AW \& NN
were involved in the LLAGN observations.

%\bibliographystyle{apj}
%\bibliography{../../Review/review}

\end{document}